\documentclass[10pt,a4paper]{article}
\usepackage[english]{babel}
\usepackage[utf8]{inputenc}
\usepackage{amsfonts,amsbsy,bm,euscript,mathrsfs}
\usepackage{amssymb,stmaryrd,faktor}
\usepackage[tbtags]{amsmath}
\usepackage{bbm}
\usepackage{float}
\usepackage{graphicx}
\usepackage[title,titletoc]{appendix}
\usepackage[bookmarks=true,colorlinks=true,linkcolor=blue,citecolor=blue,urlcolor=blue,bookmarksnumbered]{hyperref}
\usepackage{cite}

\usepackage{dsfont}
\usepackage{lmodern}
\usepackage{mathrsfs}
\usepackage{mathtools}
\usepackage{bbm}
\usepackage{braket}
\usepackage{slashed}

\usepackage{graphicx}
\usepackage{booktabs}
\usepackage{subfig}
\usepackage{tikz,tikz-feynman}
\usetikzlibrary{plotmarks,calc,decorations,decorations.pathmorphing,decorations.markings,positioning}

\numberwithin{equation}{section}

\makeatletter
\renewcommand\section{\@startsection {section}{1}{\z@}
{-3.5ex \@plus -1ex \@minus -.2ex}
{2.3ex \@plus.2ex}
{\normalfont\Large\bfseries}}
\renewcommand\subsection{\@startsection{subsection}{2}{\z@}
{-3.25ex\@plus -1ex \@minus -.2ex}
{1.5ex \@plus.2ex}
{\normalfont\large\bfseries}}
\makeatother

\newcommand{\bea}{\begin{eqnarray}}
\newcommand{\eea}{\end{eqnarray}}

\allowdisplaybreaks 

\DeclareMathOperator*{\res}{Res}

\usepackage{blkarray}

\begin{document}

\thispagestyle{empty}
\begin{flushright}\footnotesize\ttfamily
DMUS-MP-21/03
\end{flushright}
\vspace{2em}

\begin{center}

{\Large\bf \vspace{0.2cm}
{\color{black} \large Three-parameter deformation of $\mathbb{R}\times S^3$ in the Landau-Lifshitz limit}} 
\vspace{1.5cm}

\textrm{Juan Miguel Nieto García\footnote{\texttt{j.nietogarcia@surrey.ac.uk}} and Leander Wyss\footnote{\texttt{l.wyss@surrey.ac.uk}}}

\vspace{2em}

\vspace{1em}
\begingroup\itshape
Department of Mathematics, University of Surrey, Guildford, GU2 7XH, UK
\par\endgroup

\end{center}

\vspace{2em}

\begin{abstract}\noindent 
In this article we construct the effective field theory associated to the $\mathbb{R}\times S^3$ sector of the three-parameter deformation of $AdS_3 \times S^3 \times T^4$ in the Landau-Lifshitz approximation. We use this action to compute the dispersion relation of excitations around the BMN vacuum and the perturbative $S$-matrix associated to them. We are able to compute  and sum all the different loop contributions to the $S$-matrix in this limit.
\end{abstract}

\newpage

\overfullrule=0pt
\parskip=2pt
\parindent=12pt
\headheight=0.0in \headsep=0.0in \topmargin=0.0in \oddsidemargin=0in

\vspace{-3cm}
\thispagestyle{empty}
\vspace{-1cm}

\tableofcontents

\setcounter{footnote}{0}

\section{Introduction}

Integrability has had an incredible success in the construction of the spectrum of free strings propagating on backgrounds like $AdS_5 \times S^5$ or $AdS_3 \times S^3 \times M^4$ (see \cite{foundations,bigreview,Babichenko:2009dk,reviewAdS3} for reviews). This has created an interest in deformations of those backgrounds that do not spoil the integrability of the original theory. A systematic procedure to deform integrable backgrounds in such way was proposed by Klim\v{c}\'{\i}k in \cite{Klimcik:2002zj} and refined by Delduc, Magro and Vicedo in \cite{Delduc:2013fga,Delduc:2013qra}, see also \cite{Kawaguchi:2014qwa,Kawaguchi:2014fca,vanTongeren:2019dlq}. The deformations obtained through this method are usually called \emph{Yang-Baxter deformations}, due to the central place that the classical Yang-Baxter equation takes in the construction. This deformation-building technique was shown to encompass several renowned deformations of $AdS_5 \times S^5$ (like the Lunin-Maldacena and the Schrödinger backgrounds \cite{Matsumoto:2014nra,Matsumoto:2014cja,Matsumoto:2015uja,vanTongeren:2015soa}) and gave a unified understanding of them (see \cite{Borsato:2016pas,Borsato:2017qsx,Borsato:2018idb} for the connection to non-Abelian T-duality and \cite{Araujo:2017jkb,Araujo:2017jap} for the connection to $\mathcal{O}(d,d)$ transformations).

Although there exist many integrable deformations of $AdS_5 \times S^5$, the $AdS_3 \times S^3 \times T^4$ background is even richer in deformations. On the one hand, the Green-Schwarz string action of $AdS_3 \times S^3 \times T^4$, when formulated in terms of a supercoset model, has the structure of the direct product of two cosets. Thus, we are allowed to perform a different Yang-Baxter deformation to each of the factors. This two-parameter deformation is usually called a \emph{bi-Yang-Baxter deformation} \cite{Klimcik:2008eq,Klimcik:2014bta} and it was applied to the specific context of $AdS_3 \times S^3 \times T^4$ background in \cite{Hoare:2014pna,Lunin:2014tsa,Ben2}. On the other hand, both the $AdS_3 \times S^3 \times T^4$ and the $AdS_5 \times S^5$ backgrounds are supported by a Ramond-Ramond flux. However, an additional Neveu-Schwarz-Neveu-Schwarz flux can be added consistently to the $AdS_3 \times S^3 \times T^4$ background but not to the $AdS_5 \times S^5$. This \emph{mixed-flux deformation} was proven to be classically integrable \cite{Cagnazzo:2012se} and has been extensively studied from the integrability perspective \cite{Hoare:2013pma,Hoare:2013ida,Stepanchuk,Babichenko:2014yaa,Hernandez:2014eta,Lloyd:2014bsa,Pittelli:2017spf,Baggio:2018gct,Dei:2018mfl,Fontanella:2019ury,Hernandez:2019buc,Sfondrini:2020ovj,Ruiz:2021qqm}.

The mixed-flux deformation and the bi-Yang-Baxter deformation were recently combined into a \emph{three-parameter deformation} \cite{3deformedLagrangian} (see also \cite{Delduc:2014uaa,Delduc:2017fib}). However, it is still unknown if the proposed background solves the supergravity equations. This question has a non-trivial answer already in the case of the Yang-Baxter deformation of $AdS_5 \times S^5$, where it depends on the fermionic Dynkin diagram used to construct the coset model \cite{Hoare:2018ngg}. The same problem is also present in the bi-Yang-Baxter deformation of $AdS_3 \times S^3 \times T^4$ in \cite{Seibold:2019dvf}. Nevertheless, this issue does not impede the application of the usual uniform light-cone gauge quantisation procedure \cite{foundations} to construct the tree-level bosonic $S$-matrix associated to this background \cite{3deformedSMatrix}.

In this article, we will examine the $S$-matrix of the three-parameter deformation of $AdS_3 \times S^3 \times T^4$ from a different perspective. Instead of starting from the Hamiltonian associated to the theory in uniform light-cone gauge, we will construct the effective field theory associated to excitations around a BMN vacuum with large angular momentum in the large string tension limit. In the case of strings in $\mathbb{R}\times S^5 \subset AdS_5 \times S^5$ this regime provides an effective action whose leading contribution is the same as the one for the classical ferromagnet \cite{Kruczenski:2003gt,Kruczenski:2004kw,Hernandez:2004uw,Stefanski:2004cw,Kruczenski:2004cn}. Hence, this limit is usually called Landau-Lifshitz (LL) limit.\footnote{This limit has also been studied for the cases of beta \cite{Frolov:2005ty}, mixed-flux \cite{Stepanchuk}, Yang-Baxter \cite{Kentaroh,Banerjee:2017mpe} and bi-Yang-Baxter \cite{Wen:2019mgv} deformations of $AdS_3 \times S^3$ . However, these articles focus more on classical solutions and the relation between the string theory and spin chain Lagrangians. Nevertheless, the effects of the beta deformation on the perturbative quantisation were studied in \cite{Gerotto:2017sat}.} The construction can be generalised to non-compact sectors of the theory \cite{Bellucci:2004qr}, to fermionic degrees of freedom \cite{Hernandez:2004kr,Stefanski:2005tr}, and to sectors containing both \cite{Bellucci:2005vq,Bellucci:2006bv,Stefanski:2007dp}. This limit has been studied in such a great detail because it provides a direct link between the string theory action and the spin chain description of single trace operators. In addition, this setting allows for the computation of the dispersion relation of excitations around the BMN solution \cite{Minahan:2005mx,Minahan:2005qj} and the $S$-matrix associated to them \cite{Klose:2006dd,Roiban:2006yc} as a series in the string tension. Although more complete information on the tree-level $S$-matrix can be obtained from uniform light-cone gauge quantisation, this method provides easier access to information beyond tree-level. In particular, the Lagrangian associated to $\mathbb{R}\times S^3 \subset AdS_5 \times S^5$ contains enough information to reconstruct the Heisenberg model $S$-matrix and the low-energy limit of the $S$-matrix of the spin chain of Beisert, Dippel and Staudacher (BDS) \cite{Beisert:2004hm}. The factorisation of this $S$-matrix was studied up to second order in \cite{Melikyan:2008cy}.

This article is organised as follows. In section 2, we will review the geometry of the three-parameter deformation of $AdS_3 \times S^3 \times T^4$. In addition, we will construct the first two orders of its LL limit and expand them around its trivial vacuum. This vacuum corresponds to the ferromagnetic vacuum of a Heisenberg-like effective field theory, and it can be interpreted as a BMN solution of the original string sigma model. In section 3, we will perform a canonical quantisation of the Lagrangian, together with the computation of the tree-level and 1-loop corrections to the two-body $S$-matrix at leading order and next-to-leading order in the string tension. We are also able to compute the generic $n$-loop correction at tree level, which can be summed to an all-loop $S$-matrix. In addition, we will comment on a possible  extension of this result to all orders in $\lambda$ in the low energy limit. Section 4 contains some concluding remarks and future directions for this project.

\section{Generalised Landau-Lifshitz action from the three-parameter deformation of $\mathbb{R}\times S^3$}


\subsection{Non-linear string sigma model}

We will consider the bosonic non-linear sigma model action with Kalb-Ramond field given by
\begin{equation}
	S= \frac{\sqrt{\lambda}}{4\pi} \int d\tau \int d\sigma \left( \sqrt{-h} h^{ab} \partial_a X^M \partial_b X^N G_{MN} + \epsilon^{ab} \partial_a X^M \partial_b X^N B_{MN} \right) \ ,
\end{equation}
where $h^{ab}$ is the worldsheet metric and $\epsilon^{ab}$ is the Levi-Civita symbol. Here we choose the convention $\epsilon^{\tau \sigma}=1$. The background metric and the Kalb-Ramond field associated to the three-parameter deformation of $AdS_3 \times S^3$ take the form\footnote{Here we use the convention $d a\wedge d b=d a\otimes d b -d b \otimes d a$.}
\begin{align}
	ds^2&=\frac{1}{F_A}\Big[\,\frac{1-q^2 \rho^2(1+\rho^2)}{1+\rho^2}\,d \rho^2-2q \chi_{-}\rho(1+\rho^2)\, d \rho\, d t+2q \chi_+ \rho^3\, d \rho\, d \psi \notag \\
	&-\big(1+\chi_{-}^2(1+\rho^2)\big)(1+\rho^2)\,d t^2+2 \chi_+\chi_- \rho^2(1+\rho^2)\,d t\,d\psi +\rho^2 (1 - \rho^2 \chi_+^2)\, d \psi^2 \,\Big] \notag \\
	&+\frac{1}{F_S}\Big[\,\frac{1+q^2 r^2(1-r^2)}{1-r^2}\,d r^2-2q \chi_{-}r(1-r^2)\, d r\, d \phi_1-2q \chi_+ r^3\, d r\, d \phi_2 \notag \\
&+\big(1+\chi_{-}^2(1-r^2)\big)(1-r^2)\,d \phi_1^2+2 \chi_+\chi_- r^2(1-r^2)\,d\phi_1\,d\phi_2+ r^2  (1 + \chi_+^2 r^2) \,d \phi_2^2\,\Big] \ ,
\end{align}
\begin{align}
	B &=-\frac{a\,q}{F_A}\,(1+\rho^2 ) \Big[2-\rho^2 q^2-(2+\rho^2)\chi_-^2-\rho^2\chi_{+}^2 \Big]\,
d t\wedge d \psi \notag \\
	&-\frac{a\,q}{F_S}\,(1-r^2 ) \Big[2+r^2 q^2+(r^2 - 2)\chi_-^2+r^2\chi_{+}^2 \Big]\,
d \phi_1 \wedge d \phi_2 \ ,
\end{align}
where
\begin{align}
	F_{A}=&\,1-\chi_{+}^2\rho^2+\chi_-^2(1+\rho^2)-q^2\rho^2(1+\rho^2) \ ,\\
	F_{S}=&\,1+\chi_{+}^2r^2+\chi_-^2(1-r^2)+q^2r^2(1-r^2) \ , \\
	a=&\frac{1}{\sqrt{\big(q^2+\chi_+^2+\chi_-^2\big)^2+4\big(q^2-\chi_+^2\chi_-^2\big)}}\ .
\end{align}
Notice that our definition of the Kalb-Ramond field differs from both the one given in appendix C of \cite{3deformedLagrangian} and the one used by \cite{3deformedSMatrix}. The difference is a constant two-form in both cases, so it does not affect the equations of motion. However, this choice is important in the construction of classical solutions, as it affects the form of the conserved charges. Our choice is motivated by the sigma model for the $AdS_3 \times S^3$ background with mixed flux, where the dispersion relation for the dyonic giant magnon is finite only for one particular value\cite{Stepanchuk}. However, requiring our choice to match the one of the mixed-flux deformation does not unequivocally fix it. For that reason, we decided to fix this constant in such a way that the difference between our Kalb-Ramond field and the ones presented in \cite{3deformedLagrangian} and \cite{3deformedSMatrix} is independent of the three deformation parameters. This choice is further justified later, where we show that it eliminates some terms in the LL expansion.

The action presents four global isometries associated to shifts of the $t$, $\psi$, $\phi_1$ and $\phi_2$ coordinates. The conserved quantities associated to these symmetries are the space-time energy, the Lorentzian spin and two angular momenta, defined as\pagebreak
\begin{align}
	E &= \frac{\sqrt{\lambda}}{2\pi}\int_0^{2\pi} \frac{d\sigma}{F_A} \Big\{ q \chi_- \rho (1+\rho^2) \dot{\rho} - \chi_- \chi_+ \rho^2 (1+\rho^2) \dot{\psi} + [1+\chi_-^2 (1+\rho^2)](1+\rho^2) \dot{t} \notag \\
	& \qquad \qquad \qquad \qquad  -a q (1+\rho^2)[2- \rho^2 q^2 - (2+\rho^2) \chi_-^2 - \rho^2 \chi_+^2] \dot{t} \Big\} =  \sqrt{\lambda} \kappa \ , \\
	S &=\frac{\sqrt{\lambda}}{2\pi}\int_0^{2\pi} \frac{d\sigma}{F_A} \Big\{ -q \chi_+ \rho^3 \dot{\rho} -\chi_- \chi_+ \rho^2 (1+ \rho^2) \dot{t} - \rho^2 (1-\rho^2 \chi_+^2) \dot{\psi} \notag \\
	& \qquad \qquad \qquad \qquad  - a q (1+\rho^2)[2- \rho^2 q^2 - (2+\rho^2) \chi_-^2 - \rho^2 \chi_+^2] t^\prime \Big\} = \sqrt{\lambda} \mathcal{S} \ , \\
	J_1 &=\frac{\sqrt{\lambda}}{2\pi}\int_0^{2\pi} \frac{d\sigma}{F_S} \Big\{ -[1+\chi_-^2 (1-r^2)] (1-r^2) \dot{\phi_1} - \chi_+ \chi_- r^2 (1-r^2) \dot{\phi_2} + q \chi_- r (1-r^2) \dot{r} \notag \\
	& \qquad \qquad \qquad \qquad - a q (1-r^2) [2+r^2 q^2 + (r^2-2) \chi_-^2 + r^2 \chi_+^2] \phi_2^\prime \Big\} = \sqrt{\lambda} \mathcal{J}_1 \ , \\
	J_2 &=\frac{\sqrt{\lambda}}{2\pi}\int_0^{2\pi} \frac{d\sigma}{F_S} \Big\{ -r^2 (1+\chi_+^2 r^2) \dot{\phi_2} - \chi_+ \chi_- r^2 (1-r^2) \dot{\phi_1} + q \chi_+ r^3 \dot{r} \notag \\
	& \qquad \qquad \qquad \qquad + a q (1-r^2) [2+r^2 q^2 + (r^2-2) \chi_-^2 + r^2 \chi_+^2] \phi_1^\prime \Big\} = \sqrt{\lambda} \mathcal{J}_2  \ ,
\end{align}
where prime denotes derivatives with respect to $\sigma$ and dot denotes derivatives with respect to $\tau$.

Here we will be interested only on the $\mathbb{R}\times S^3\subset AdS_3 \times S^3$ subspace defined by setting $\rho=0$, which simplifies the first two lines of the metric to just $-d t^2$. In addition, this fixes $S=0$.


\subsection{Generalised classical LL action}

The construction of the LL action for our case follows the same steps as the original $AdS_5\times S^5$ construction: we consider a rotating string with large semiclassical angular momentum, isolate the ``fast'' coordinate, gauge fix it and construct the effective Lagrangian for the ``slow'' coordinate. In our case, the regime of large total angular momentum $J=J_1 + J_2$ means that we have to consider the sum $\phi_1 + \phi_2$ as our fast coordinate.

The LL effective field theory is usually constructed as the large $\mathcal{J}=\mathcal{J}_1+\mathcal{J}_2$ expansion of the light-cone gauge Hamiltonian. However, because we will only need the two leading orders in $\mathcal{J}$ for our purposes, we can use a different gauge that simplifies the computations. We will fix the world-sheet metric to the standard conformal gauge $h^{ab}= \text{diag}(-1,1)$ and fix the residual conformal symmetry by picking the condition $t=\kappa \tau$. Instead of an expansion in inverse powers of the angular momentum, we get an expansion in inverse powers of the energy $\kappa$. The key observation is that these two quantities are equal at leading order
\begin{equation}
	\frac{1}{\mathcal{J}^2}=\frac{1}{\kappa^2} + \mathcal{O} \left( \frac{1}{\kappa^4} \right) \ ,
\end{equation}
meaning that the two first orders of both expansions agree \cite{Kruczenski:2004kw}.

The first step in our computation is to perform a change of coordinates that isolates the fast coordinate
\begin{equation}
	\phi_1 = \alpha + \beta + \kappa \tau \ , \qquad \phi_2=\alpha - \beta + \kappa \tau \ .
\end{equation}
This way we decouple the fast behaviour and introduce the two ``slow'' coordinates $\alpha$ and $\beta$. In these coordinates, the Lagrangian takes the form
\begin{align}
	&\frac{4\pi}{\sqrt{\lambda}} F_S \mathcal{L} = \frac{\left[q^2 r^2 \left(r^2-1\right)-1\right]
   \left(\dot{r}^2-{r^\prime}^2 \right)}{r^2-1} + \left(1-r^2\right) \left[1+\chi_-^2 \left(1-r^2\right)\right] \left[\left(\kappa +\dot{\alpha}+\dot{\beta}\right)^2-\left(\alpha^\prime+\beta^\prime \right)^2\right] \notag \\
	&-2 \chi_- \chi_+ \left(r^2-1\right) r^2 \left[(\kappa +{\dot{\alpha}})^2-{\dot{\beta}}^2-{\alpha^\prime}^2+{\beta^\prime}^2\right]+r^2 \left(1+\chi_+^2 r^2\right) \left[\left(\kappa +\dot{\alpha}-\dot{\beta}\right)^2-\left(\alpha^\prime-\beta^\prime\right)^2\right]\notag  \\	
	&-2 a q \left(1-r^2\right) \left[r^2 \left(\chi_+^2-\chi_-^2+q^2\right)+2 \chi_-^2+2\right] \left[\beta^\prime \left(\kappa +\dot{\alpha}\right)-\dot{\beta} \alpha^\prime \right]-\kappa^2 F_S \ . \label{beforeexpansion}
\end{align}

Before expanding the action in inverse powers of $\kappa$, we have to make sure that the dynamic of the string is mainly captured by the fast behaviour we have just decoupled. One way to do so is rescaling the time coordinate as $\tau \rightarrow \kappa\tau$ and $\partial_\tau\rightarrow \partial_\tau/\kappa$ (which is equivalent to writing the time derivatives in terms of $t$ instead of $\tau$), so terms quadratic in the generalised velocities do not appear in the leading order of our expansion. Notice that this changes the prefactor in the action from $\frac{\sqrt{\lambda}}{4\pi}$ to $\frac{\sqrt{\lambda} \kappa}{4\pi}\approx \frac{J}{4\pi}$.

However, if we take the large $\kappa$ limit straightforwardly, the leading contribution to our action will be
\begin{equation}
	\frac{(\chi_+ - \chi_-)^2 +q^2}{F_S} r^2 (1-r^2) \kappa^2 \ .
\end{equation}
This term contains no derivatives, so it carries no dynamics. Derivatives appear first at order $\kappa^0$ on our Lagrangian, so we would like to make the above term of order $\kappa^0$ instead of $\kappa^2$. This can be attained by rescaling all the deformation parameters with a factor of $\kappa^{-1}$, but this naïve action freezes the constant in front of the flux term to $aq=\frac{1}{2}$. If we want the flux term to appear at leading order the action with a tunable parameter, we need to perform the rescaling
\begin{equation}
	\chi_\pm \rightarrow \chi_\pm/ \kappa \ , \qquad q \rightarrow q/\kappa^3 \ .
\end{equation}
Then, the leading order of the action takes the form
\begin{equation}
	\mathcal{L}^{(0)}= - 2 [\dot{\alpha} + (1-2 r^2) \dot{\beta} ] + \frac{r^{\prime 2}}{1-r^2} + \alpha^{\prime 2} + \beta^{\prime 2}+2 (1-2r^2) \alpha ' \beta ' + 4 \tilde{q} (1-r^2) \beta' -(\chi_+ - \chi_-)^2 r^2 (1-r^2) +\mathcal{O} \left( \frac{1}{\kappa^2} \right) \ ,
\end{equation}
where $\tilde{q} \sqrt{(\chi_+^2-\chi_-^2)^2}=q$. Notice that the term $-2\dot{\alpha}$ is a total derivative and can be immediately discarded.

In addition to the equations of motion from the Lagrangian, the dynamics of the string are constrained by the Virasoro constraints. If we perform the same rescaling and expansion to the Virasoro constraints, we obtain
\begin{gather}
\label{vivi}
\frac{r^{\prime 2}}{1-r^2} -(\chi_+ - \chi_-)^2 r^2 (1- r^2) + 2 [\dot{\alpha} +(1 - 2 r^2) \dot{\beta} ] + (1 - 2 r^2) [ \alpha^{\prime 2} +\beta^{\prime 2}] + 2 \alpha' \beta'=\mathcal{O} \left( \frac{1}{\kappa^2} \right) \ , \\
[\alpha' + (1-2 r^2) \beta' ] \kappa =\mathcal{O} \left( \frac{1}{\kappa} \right) \ .
\end{gather}
The second equation allows us to completely eliminate the variable $\alpha$ from the Lagrangian, giving us\footnote{Usually substituting the Virasoro constraints directly in the action gives rise to the wrong Lagrangian. This proves not to be a problem in the leading order of our Lagrangian. We have checked that the equations of motion obtained by substituting the Virasoro constraints at the level of the Lagrangian are the same as those obtained by substituting such constraints at the level of the equations of motion. The same behaviour has been observed in the undeformed case \cite{Kruczenski:2004kw} and in the mixed-flux deformation \cite{Stepanchuk}.}
\begin{equation}
	\mathcal{L}^{(0)}=- 2 (1-2 r^2) \dot{\beta}  - 4 r^2 (1-r^2) \beta^{\prime 2}+ 4 \tilde{q} (1-r^2) \beta' + \frac{r^{\prime 2}}{1-r^2}-(\chi_+ - \chi_-)^2 r^2 (1-r^2)  +\mathcal{O} \left( \frac{1}{\kappa^2} \right) \ .
\end{equation}
If we turn off the flux and the deformation $\chi_-$, this Lagrangian reduces to the LL effective Lagrangian for strings in $\eta$-deformed $\mathbb{R}\times S^3$ computed in \cite{Kentaroh}. If we instead turn off both deformation parameters $\chi_\pm$, this Lagrangian reduces to the LL effective Lagrangian for strings in $\mathbb{R}\times S^3$ supported by a mixture of R-R flux and NS-NS flux \cite{Stepanchuk}.

As a last step, it is more convenient to combine the fields $r$ and $\beta$ into a single complex scalar. This can be done by defining the field $\phi=\sqrt{1-r^2} e^{2 i \beta}$\pagebreak
\begin{align}
	\mathcal{L}^{(0)}&=\frac{i (1-2 |\phi |^2 ) \phi^* \dot{\phi}}{2 |\phi |^2 }+i \tilde{q} \phi^*  \phi'-\frac{i (1-2 |\phi |^2 ) \phi \dot{\phi}^*}{2 |\phi |^2 }-i \tilde{q} \phi  \phi^{* \prime}-(\chi_+ -\chi_-)^2 |\phi |^2  (1-|\phi |^2 ) \notag \\
	&-\frac{1}{2} (1-|\phi |^2 ) \phi' \phi^{* \prime}-\frac{(\phi^*)^2 (2-|\phi |^2 ) \phi'^2+2 \phi' \phi^{* \prime}+\phi ^2 (2-|\phi |^2 ) (\phi^{* \prime})^2}{4 (1-|\phi |^2 )} \ .
\end{align}
One may be worried by the imaginary unit appearing in the action, but this is not a problem, as they always appear accompanying a derivative.

In the next section we will be interested in computing the $S$-matrix associated to this Lagrangian. For this purpose, we should not consider just the limit of large $\kappa$, but the large $J$ and large $\lambda$ limits independently, together with a rescaling that decompactifies the spatial coordinate $\sigma$. This last rescaling should be accompanied by a similar rescaling of the deformation parameters
\begin{equation}
	t \rightarrow t/J^2 \ , \qquad\sigma \rightarrow \sigma/J \ , \qquad \chi_\pm \rightarrow J\chi_\pm \ , \qquad \tilde{q} \rightarrow J\tilde{q} \ . \label{secondscaling}
\end{equation} 
Notice that we are still in the small deformation regime, as the total rescaling of the deformation parameters is $\chi_\pm \rightarrow \chi_\pm /\sqrt{\lambda}$, $q\rightarrow q/\sqrt{\lambda^3}$. In addition to the above rescaling, we will perform an expansion of our fields around $\phi=0$. This corresponds to computing fluctuations around the BMN solutions. To do so, we first have to rescale the field $\phi$ as $\phi/\sqrt{J}$. We can do so because the factor $J$ in front of the LL action plays the rôle of the inverse Planck constant \cite{Minahan:2005qj}. Then, the terms quadratic and quartic in the fields take the form\footnote{Here we have included an extra factor $-2\dot{\beta}$ into the action before performing the limit. We have done so to get rid of a term of order $J^{2}$ that would have appeared otherwise. Because this term is a total derivative, it does not affect the equations of motion. In addition, another term of order $J^{2}$ appears if we add a constant to the Kalb-Ramond field. Although this term vanishes for our choice, it is also a total derivative and it is irrelevant for the equations of motion.}
\begin{align}
	J^{-1} \mathcal{L}^{(0)}&=\phi^* (-i\partial_t +i \tilde{q} \partial_\sigma) \phi - \phi (-i\partial_t +i \tilde{q} \partial_\sigma) \phi^*-(\chi_+ - \chi_-)^2 \phi^* \phi -\phi^{* \prime} \phi' \notag \\
	&-\frac{1}{2J} \left[(\phi^*)^2 \phi^{\prime 2} + \phi^2(\phi^{*\prime}) ^2 - 2 (\chi_- - \chi_+)^2 \phi^2 (\phi^*)^2 \right] + \mathcal{O} \left( \frac{1}{J^2} \right) \ . \label{LeadingL}
\end{align}

A similar process can be performed to the order $\kappa^{-2}$ of the expansion, which becomes $\lambda^{-1}$ after the last rescaling. We have collected the details in appendix~\ref{subleading} due to the length of the expressions involved. The final result being
\begin{align}
	\mathcal{L} &=\frac{\sqrt{\lambda} \kappa}{4\pi} \left[\mathcal{L}^{(0)}+\lambda^{-1} \mathcal{L}^{(2)} + \lambda^{-2} \mathcal{L}^{(4)} +\dots \right]  \ , \label{expandedlagrangian} \\
	J^{-1} \mathcal{L}^{(2)} &=\frac{1}{4} \Big[ (\phi^{* \prime \prime} \phi'' + (\chi_-^2 - \chi_+^2)^2 |\phi |^2 - (\chi_-^2 - \chi_+^2)  (\phi^{* \prime \prime} \phi + \phi^* \phi'') - 2i\tilde{q} ( \phi^{* \prime \prime} \phi' + \phi^{* \prime} \phi'') \notag \\
	&+ 4 (\chi_+^2 + \tilde{q}^2) \phi^{* \prime} \phi' + 8 i \tilde{q}^3 (\phi^{* \prime}  \phi - \phi^* \phi') \Big]+\frac{1}{8 J} \Big[ 2 (\phi^{* \prime})^2 \phi'^2 -  2 \phi^* \phi^{* \prime \prime} \phi'^2 -2 (\phi^{* \prime})^2 \phi \phi'' \notag \\
	&+4 \phi^{* \prime}\phi' \left( \phi^* \phi'' + \phi^{* \prime \prime} \phi \right) +(\phi^*)^2 \phi''^2 + (\phi^{* \prime \prime})^2  \phi^2 + (\chi_-^2 - \chi_- \chi_+ + \chi_+^2) [(\phi^*)^2 \phi'^2 + (\phi^{* \prime})^2 \phi^2 ] \notag \\
	&-8 \chi_- \chi_+ (\chi_- - \chi_+)^2 |\phi|^4 -4i \tilde{q} ( \phi^{* \prime} \phi - \phi^* \phi') [ (\chi_- - \chi_+)^2 |\phi|^2 +2 \phi^{* \prime} \phi' ] +4 \tilde{q}^2 [ (\phi^*)^2 \phi'^2  \notag \\
	&+ (\phi^{* \prime})^2 \phi^2] \Big]+ \mathcal{O} \left( \frac{1}{J^2} \right) \ . \label{SecondL}
\end{align}

We will not consider higher order terms in $1/J$ as they are irrelevant for our computation. We will show in the next section that the two-body $S$-matrix is completely described only by the vertex associated to the term quartic in fields.

\section{Quantisation and $S$-matrix}

In the previous section we have constructed a Lagrangian that describes the fluctuations around the BMN vacuum in the limit of large angular momentum, $J$, and large string tension, $\lambda$. This action resembles the action for the positive-energy part of a massive relativistic complex scalar field, or a modified version of the Lagrangian associated to a non-linear Schrödinger equation. In this section we will canonically quantise the quadratic part of the Lagrangian and use the quartic term of the Lagrangian to compute the two-body $S$-matrix. We will compare some of our results with the dispersion relation and the $S$-matrix computed in \cite{3deformedSMatrix}.

\subsection{Quantisation near the BPS vacuum}

The quadratic part of the action we computed in the previous section is given by
\begin{align}
\mathcal{L}_2&=\phi^* (-i\partial_t +i \tilde{q} \partial_\sigma) \phi - \phi (-i\partial_t +i \tilde{q} \partial_\sigma) \phi^*-(\chi_+ - \chi_-)^2 \phi^* \phi -\phi^{* \prime} \phi' + \notag \\
&+\frac{1}{4\lambda} \Big[ (\phi^{* \prime \prime} \phi'' + (\chi_-^2 - \chi_+^2)^2 |\phi |^2 - (\chi_-^2 - \chi_+^2)  (\phi^{* \prime \prime} \phi + \phi^* \phi'') - 2i\tilde{q} ( \phi^{* \prime \prime} \phi' + \phi^{* \prime} \phi'') \notag \\
&+ 4 (\chi_+^2 + \tilde{q}^2) \phi^{* \prime} \phi' + 8 i \tilde{q}^3 (\phi^{* \prime}  \phi - \phi^* \phi') \Big] +\mathcal{O} \left( \frac{1}{\lambda^2} \right) \ . \label{quadraticlagrangian}
\end{align}
The most important characteristic of this Lagrangian is that generalised velocities only appear linearly. This means that, in contrast with the usual relativistic QFT quantisation, we can construct the field operator in the interaction picture using only negative-frequency modes
\begin{equation}
	\phi (t,\sigma)=\int{\frac{dp}{\sqrt{2\pi \omega(p)}} a_p e^{-i\omega(p) t + ip\sigma}} \ , \qquad \phi^* (t,\sigma)=\int{\frac{dp}{\sqrt{2\pi \omega(p)}} a_p^\dagger e^{i\omega(p) t - ip\sigma} }\ ,
\end{equation}
where $a_p$ and $a^\dagger_p$ are creation and annihilation operators respectively. As our Lagrangian is canonically normalised after multiplying it by a factor of $-\frac{1}{2}$, they satisfy usual commutation relations, $[a_p , a^\dagger_k] =(2 \pi) \delta (p-k)$. In addition, this also implies that the ground state can be chosen such that it is annihilated by the field operator
\begin{equation}
	\phi (t,\sigma) |0\rangle =0 \ .
\end{equation}

The function $\omega(p)$ appearing in the plane wave expansion of the field operator is fixed by the equations of motion. For the Lagrangian (\ref{quadraticlagrangian}) we have
\begin{equation}
	\omega (p)=\frac{(\chi_- -\chi_+)^2+p (p-2\tilde{q})}{2} -\frac{(\chi_-^2 - \chi_+^2)^2 +p^2 (p-2\tilde{q})^2 +2 (\chi_-^2 + \chi_+^2) p^2 -16\tilde{q}^3 p}{8\lambda} + \mathcal{O} \left( \frac{1}{\lambda^2} \right) \ . \label{dispersion}
\end{equation}

This result can be checked against the expression computed in \cite{3deformedSMatrix} from the uniform light-cone gauge quantisation of the non-linear string sigma model
\begin{equation}
	\Omega_\pm (p)=\sqrt{p^2 \pm 2 a q p (2+q^2+\chi_-^2 + \chi_+^2) +q^2 +(1+\chi_+^2)(1+\chi_-^2)} \pm \chi_+ \chi_- \ , \label{ExactEnergy}
\end{equation}
where the $\pm$ depends on the field considered. In order to compare their result with ours, we need to  consider the low-momentum and small-deformation limit of this expression. This can be done by rescaling the involved quantities with the string tension $\lambda$ in the following way $p\rightarrow p/\sqrt{\lambda}$, $\chi_\pm \rightarrow \chi_\pm /\sqrt{\lambda}$, $q\rightarrow q/\sqrt{\lambda^3}$ and taking $\lambda\rightarrow \infty$. By doing so, we find a perfect match with the two first non-trivial orders
\begin{equation}
	\Omega_- (p/\lambda) \approx 1+\frac{\omega (p)}{\lambda} + \mathcal{O} \left( \frac{1}{\lambda^3} \right) \ , \label{energyrelation}
\end{equation}
where the factor $1$ comes from the subtraction of the fast coordinate.\footnote{One might have thought that the restriction to small deformation would have spoiled this relation. This does not happen because the light-cone gauge quantisation assume implicitly that the deformation parameters are small with respect to the string tension.}

Another consequence of having a Lagrangian where generalised velocities appear only linearly is that the momentum representation of the propagator has only one pole. Of the two possible ways we can implement the $i\epsilon$ prescription, we chose the retarded option\footnote{This can also be understood from the fact that our ground state is annihilated by the field operator. As a consequence, the excitations we are creating cannot travel backwards in time.}
\begin{equation}
	D(t,\sigma)=\langle 0 | T \left\{ \phi(t,\sigma) \phi^*(0,0) \right\} |0\rangle= \int{\frac{d\Omega dp}{2\pi} \frac{i}{\Omega - \omega(p) +i\epsilon} e^{-i\Omega t +ip\sigma}}=\int{\frac{d\Omega dp}{2\pi} D(\Omega,p) e^{-i\Omega t +ip\sigma}} \ .
\end{equation}
The consequences of this choice are discussed at length in \cite{Klose:2006dd}. Among them, we want to highlight that neither the energy of the vacuum state nor the one particle Green's function are renormalised, and that the two-body $S$-matrix is given only by the sum of bubble diagrams. The first one reflects the BPS protection of the vacuum we are expanding around. The second one implies that the dispersion relation $\omega(p)$ we computed above does not receive any quantum corrections. The last one arises from the fact that, if we cut a generic diagram, the number of propagators always has to be the same and equal to the number of external incoming legs. This happens because there cannot be any past-directed propagator together with the charge conservation from the $U(1)$ symmetry of the Lagrangian. Thus, only bubble diagrams can contribute to the two-body $S$-matrix. Notice that the bubble diagrams only contain quartic vertices. Because the number of fields appearing in our Lagrangian~(\ref{expandedlagrangian}) is controlled by the expansion on $1/J$, this means that the two leading contributions in $1/J$ are enough to compute the two-body $S$-matrix.

\subsection{Field theory $S$-matrix}

In this section we will compute the two-body $S$-matrix associated to (\ref{expandedlagrangian}). As we commented in above, the only diagrams that contribute are diagrams with $n$ bubbles (we will consider the tree level contribution as $n=0$). First, we define the two-particles in-state and out-state as
\begin{equation}
	|p , p' \rangle=a^\dagger (p) a^\dagger (p') |0 \rangle \ , \qquad \langle k , k'|=\langle 0 | a(k') a(k) \ .
\end{equation}
Thus, the $S$-matrix we are interested in can be defined as
\begin{equation}
	S(p,p',k,k')=\langle k , k' | T e^{-i \int{dt d\sigma V_4}} | p , p'\rangle \ , \label{Smatrix}
\end{equation}
where $V_4$ is the contribution in (\ref{expandedlagrangian}) that is quartic in fields. As we have written the Lagrangian as a series in $\lambda$, we can expand the vertex as
\begin{equation}
	V_4=V^{(0)}_4 + \frac{V^{(1)}_4}{\lambda} + \mathcal{O} (\lambda^{-2}) \ .
\end{equation}
Thus, the leading order contribution to the $S$-matrix will be given by bubble diagrams where all the vertices are of type $V^{(0)}_4$, while the next-to-leading order in $\lambda$ will be given by diagrams with one vertex of type $V^{(1)}_4$ while the remaining vertices are of type $V^{(0)}_4$, and so on.

In addition, as our action is invariant under $t$ and $\sigma$ translations, the integration over these variables will contribute with a conservation of the energy and momentum at every vertex. Thus, these two quantities will be conserved throughout all the process, and the $S$-matrix has to be proportional to
\begin{align}
	\delta [\omega (p) + \omega(p') -\omega (k) - \omega(k')] \delta(p+p'-k-k')=\frac{\delta(p-k) \delta(p'-k') + \delta(p-k') \delta(p'-k)}{\frac{d\omega(p)}{d p} - \frac{d \omega(p')}{d p'}} \ .
\end{align}
If we substitute our expression for the dispersion relation (\ref{dispersion}), the kinematic factor accompanying the delta functions takes the form
\begin{equation}
	K(p,p')=\frac{1}{\frac{d\omega(p)}{dp} - \frac{\omega(p')}{dp'}}=\frac{1}{p-p'}+\frac{p^2+pp'+p^{\prime 2} +\chi_-^2 + \chi_+^2 -3 \tilde{q} (p+p') +2\tilde{q}^2}{2(p-p')\lambda} + \mathcal{O} \left( \frac{1}{\lambda^2} \right) \ .
\end{equation}

\subsubsection{Tree-level contribution at leading order in $\lambda$}

From of the leading order term in $\lambda$ of our Lagrangian (\ref{LeadingL}), we get the following contribution to the quartic vertex\footnote{Remember that we have to multiply the Lagrangian by $-\frac{1}{2}$ to canonically normalise the fields. In addition, we should take into account that we are working with a complex scalar, so the symmetry factor for standard diagrams is $(2!)^2$. We will also ignore the $1/J$ factor, as it can be reinstated at the end of the computations by the substitution $i\rightarrow i/J$.}
\begin{equation}
	-\langle k , k' | V_4^{(0)} | p , p'\rangle=-V^{(0)}_4(p,p',k,k') =2(\chi_- - \chi_+)^2 +p p' + k k' \ .
\end{equation}
If we incorporate the kinematic factor from the delta functions, the final result is
\begin{equation}
	S_{\text{tree}}^{(0)}=-i V^{(0)}_4(p,p',p,p') K(p,p')=2i\frac{p p' +(\chi_- - \chi_+)^2}{p-p'} \ ,
\end{equation}
where we have used that $V^{(0)}_4(p,p',p,p')=V^{(0)}_4(p,p',p',p)$.

We can compare this result with the $S$-matrix element $S_{YY}$ computed in \cite{3deformedSMatrix}. To perform this comparison we have to first rescale the momentum and deformation parameters and then take the limit of large tension, as we did to compare with their dispersion relation. The resulting expression
\begin{equation}
	S_{YY}\approx i \left[\frac{ 4(\chi_- - \chi_+)^2 +(p + p')^2}{2 (p-p')} + i(\alpha- \frac{1}{2}) (p-p') \right] \ ,
\end{equation}
where $\alpha$ is their light-cone gauge fixing parameter. We can see that this expression matches ours when we set $\alpha=-\frac{1}{2}$.

In addition, we can check that our result also matches with equation (5.26) from \cite{Gerotto:2017sat} when we set $\beta_R=\Delta=0$ and $\beta_I=(\chi_+ -\chi_-)$, although they do not consider any contribution from the flux.

\subsubsection{1-loop contribution at leading order in $\lambda$ and sum over bubbles}

Let us now consider the 1-loop correction to the $S$-matrix. Thanks to the energy and momentum conservation, we only have to compute the bubble diagram with $k=p$ and $k'=p'$
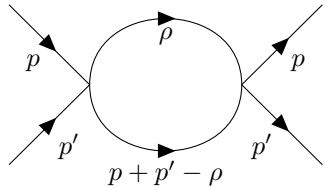
\begin{figure}[H]
\center
\begin{tikzpicture}[baseline=(current  bounding  box.center)]
\begin{feynman}
\vertex (x);
\vertex[right=2cm of x] (y);
\vertex[above left=of x] (a);
\vertex[below left=of x] (b);
\vertex[above right=of y] (c);
\vertex[below right=of y] (d);
\vertex[above right= 0.5cm and 0.75cm of x] (p);

\diagram*{
    (x) --[fermion, half left,edge label'={$\rho$}] (y),
    (x) --[fermion, half right,edge label'={$p+p'-\rho$}] (y),
    (a) --[fermion,edge label'={$p$}] (x),
    (b) --[fermion,edge label'={$p'$}] (x),
    (y) --[fermion,edge label'={$p$}] (c),
    (y) --[fermion,edge label'={$p'$}] (d),
};
\end{feynman}
\end{tikzpicture}
\end{figure}
\noindent which is given by the integral
\begin{align}
	I^{(1)}&=\int{\frac{d\Omega d\rho}{2\pi} D(\Omega,\rho) D[\omega(p) + \omega(p') -\Omega -\omega(\rho),p+p'-\rho] \left[ V_4^{(0)} (p,p',\rho,p+p'-\rho) \right]^2}= \notag \\
	&= \int{\frac{d\Omega d\rho}{2\pi} \frac{-[2(\chi_- - \chi_+)^2 +p p' + \rho (p+p'-\rho)]^2}{[\Omega - \omega(\rho) +i\epsilon][\omega(p) + \omega(p') -\Omega -\omega(p+p'-\rho) +i\epsilon]}} \ .
\end{align}
The integral over the energy of the virtual particle with momentum $\rho$ can be performed by contour integration. The integrand has two poles, one in the lower half plane from $D(\omega,\rho)$ and another in the upper half plane from the other propagator. Both ways of closing the contour give us
\begin{equation}
	I^{(1)}=\int{d\rho \frac{-[2(\chi_- - \chi_+)^2 +p p' + \rho (p+p'-\rho)]^2}{\omega(p) + \omega(p') -\omega(\rho) -\omega(p+p'-\rho) +2 i\epsilon}} \ .
\end{equation}
%

\noindent Naïve  power counting shows that the momentum integral is divergent. In order to evaluate it, we will perform contour integration with a sharp cut-off. Substituting the leading order contribution to the energy and closing the contour in the upper half plane (assuming $p>p'$), we get
\begin{align}
	I^{(1)}&\approx \int{d\rho \frac{-[2(\chi_- - \chi_+)^2 +p p' + \rho (p+p'-\rho)]^2}{p(p-2\tilde{q})+p'(p'-2\tilde{q})-\rho(\rho-2\tilde{q})-(p+p'-\rho)(p+p'-\rho-2\tilde{q}) +\mathcal{O} (\lambda^{-1}) +2 i\epsilon}} \notag \\
	&=\int{d\rho \frac{-[2(\chi_- - \chi_+)^2 +p p' + \rho (p+p'-\rho)]^2}{p^2 +p^{\prime 2} -\rho^2 -(p+p'-\rho)^2  +\mathcal{O} (\lambda^{-1}) +2 i\epsilon}}= \notag \\
	&=\res_{\rho\rightarrow p} \frac{-[2(\chi_- - \chi_+)^2 +p p' + \rho (p+p'-\rho)]^2}{p^2 +p^{\prime 2} -\rho^2 -(p+p'-\rho)^2  +\mathcal{O} (\lambda^{-1}) +2 i\epsilon} \notag \\
	&- \int_0^\pi{\Lambda e^{i\theta} d\theta \frac{-[2(\chi_- - \chi_+)^2 +p p' + \Lambda e^{i\theta} (p+p'-\Lambda e^{i\theta})]^2}{p^2 +p^{\prime 2} -\Lambda^2 e^{2i\theta} -(p+p'-\Lambda e^{i\theta})^2  +\mathcal{O} (\lambda^{-1}) +2 i\epsilon}} \ .
\end{align}
Expanding the last integral in powers of $\Lambda$, we find
\begin{align}
	&\int_0^\pi{\Lambda e^{i\theta} d\theta \frac{-[2(\chi_- - \chi_+)^2 +p p' + \Lambda e^{i\theta} (p+p'-\Lambda e^{i\theta})]^2}{p^2 +p^{\prime 2} -\Lambda^2 e^{2i\theta} -(p+p'-\Lambda e^{i\theta})^2  +\mathcal{O} (\lambda^{-1}) +2 i\epsilon}} \approx \int_0^\pi{ \Lambda e^{i\theta} d\theta \left[\frac{\Lambda^2 e^{2i\theta}}{2}+\frac{(p+p')}{2} \Lambda e^{i \theta}  \right.} \notag \\
	&\left.+2(\chi_- - \chi_+)^2 +p p'-\frac{(p+p')^2}{4}+\frac{p^2 +p^{\prime 2}}{4} +0\cdot \Lambda^{-1} +\mathcal{O} (\Lambda^{-2}) \right] \ .
\end{align}
We will use here the dimensional regularisation prescription $\int d\rho \rho^\alpha=0$ for $\alpha\geq 0$, which is equivalent to dropping all positive powers of $\Lambda$ in the integral. The final expression for the leading order in $\lambda$ of the integral is
\begin{equation}
	I^{(1)}=(p-p') \left( \frac{2(\chi_- - \chi_+)^2 +2 p p'}{p-p'} \right)^2 \ .
\end{equation}
Adding the $(-i)^2$ factor, the kinematic factor from the delta function and a factor $\frac{1}{2}$ from symmetries of the Feynman diagram, we get the 1-loop correction to the $S$-matrix
\begin{equation}
	S_{\text{1-loop}}^{(0)}=\frac{(-i)^2}{2} K(p,p') I^{(1)}= -2 \left( \frac{(\chi_- - \chi_+)^2+p p'}{p-p'} \right)^2 \ .
\end{equation}

This computation can be extended easily to the diagram with $n$ bubbles, as each bubble contributes with the same factor, giving us
\begin{equation}
	S_{\text{n-loop}}^{(0)}= 2 i^n \left( \frac{(\chi_- - \chi_+)^2+ p p'}{p-p'} \right)^n \ .
\end{equation}
Combining all our results, we can sum all the loop contributions to the $S$-matrix
\begin{align}
 S^{(0)}(p,p') &=1+S_{\text{tree}}^{(0)}+\sum_{n=1}^\infty S_{\text{n-loop}}^{(0)}= 1+2 \sum_{n=0}^\infty i^n \left( \frac{(\chi_- - \chi_+)^2+p p'}{p-p'} \right)^n \notag \\
 &=\frac{p -p' +i[(\chi_- - \chi_+)^2+p p']}{p -p' -i[(\chi_- - \chi_+)^2 +p p']}=\frac{\frac{1}{p} -\frac{1}{p'} -i\frac{(\chi_- - \chi_+)^2+p p'}{p p'}}{\frac{1}{p} -\frac{1}{p'} +i\frac{(\chi_- - \chi_+)^2 +p p'}{p p'}} \ .
\end{align}
We should point out three characteristics of this $S$-matrix. First, it is independent of the flux parameter $\tilde{q}$. We believe this might be just an artefact of our rescaling of the deformation parameters. Second, it fulfils physical and braiding unitarity if the deformation parameters are real, which means that it is a pure phase. And third, it reduces to the usual Heisenberg $S$-matrix when we turn off the deformation parameters  upon the identification $u=\frac{1}{p}$.

We may consider this $S$-matrix incomplete because our Lagrangian only describes one of the four possible excitations around the BMN vacuum considered in \cite{3deformedSMatrix}. In addition, we are also considering neither the contributions from the $T^4$ coordinates nor from fermions. Despite not taking into account these excitations in our Lagrangian, we can prove that contributions from diagrams involving them inside loops do not modify the above two-body $S$-matrix. The argument for our case follows closely the one laid down in \cite{Roiban:2006yc} for the $SU(2)$ sector of undeformed $AdS_5\times S^5$. As each type of excitation is associated to a node of a Dynkin diagram, there exists an abelian conserved charge associated to them. This means that the vertices associated to a Lagrangian that describes all possible excitations have to conserve the quantum numbers associated to those charges. This has two consequences: first, that there are no cubic vertices, and second, that there are no vertices that change two excitations of one type into two excitations of another type. In addition, because all excitations in this Lagrangian are magnons around the ferromagnetic ground state, the propagator associated to each of them can be chosen to be retarded. We conclude from these two details that loop contributions containing different excitations than those on the external legs must vanish, either because they would need vertices that do not conserve the quantum number or because the loop cannot be constructed with two retarded propagators.  This means that the $S$-matrix we obtained is complete within the context of the expansions we are considering.

\subsubsection{Tree-level and $1$-loop contribution at next-to-leading order in $\lambda$}

We can now use the next-to-leading term in the Lagrangian (\ref{SecondL}) to compute the first correction in $\lambda$ of the $S$-matrix. There exist two contributions to this quantity, one coming from the next-to-leading contribution of the vertex and another coming from the next-to-leading contribution of the kinematic factor accompanying the delta function.

We can compute the next-to-leading order contribution to the vertex from~(\ref{SecondL}), giving us
\begin{align}
	&4\lambda \langle k , k' | V_4^{(1)} | p , p'\rangle =4\lambda V^{(1)}_4(p,p',k,k') =8\chi_+ \chi_- (\chi_- - \chi_+)^2 +p p' k (p+p'+k+4\tilde{q}) \notag\\
	&+p p' k' (p+p'+k'+4\tilde{q})+p k k' (p+k+k'+4\tilde{q})+p' k k' (p'+k+k'+4\tilde{q}) \notag \\
	&+2\tilde{q} (\chi_- - \chi_+)^2 (p + p' +k + k')-(p p'+k k' )\left[p p'+k k' -4\chi_+ \chi_- -4(\chi_- - \chi_+)^2 -4\tilde{q}^2 \right] \ . \label{subleadingvertex}
\end{align}

Thus, the tree-level $S$-matrix at order $\lambda^{-1}$ is given by
\begin{align}
	&\lambda S_{\text{tree}}^{(1)}=-i \left[ K^{(0)} V^{(1)}_4(p,p',p,p') + K^{(1)} V^{(0)}_4(p,p',p,p') \right] \notag \\
	&=-i\frac{2 (\chi_+ \chi_- + p p') (\chi_- - \chi_+)^2 + p p' (p^2+p^{\prime 2}) + \tilde{q} (p+p') [2p p' + (\chi_- - \chi_+)^2] +2 p p'\left[\chi_+ \chi_- +\tilde{q}^2 \right]}{p-p'} \notag \\
	&+i\frac{[p p' +(\chi_- - \chi_+)^2][p^2+pp'+p^{\prime 2} +\chi_-^2 + \chi_+^2 -3 \tilde{q} (p+p') +2\tilde{q}^2]}{(p-p')^2} \ .
\end{align}

For the case of the $1$-loop level, we have three different contributions at the next-to-leading order: from the kinematic factor, from the next-to-leading contribution to the vertex and from the next-to-leading contribution to the dispersion relation.

The first contribution can be written with the information we already have at our disposal
\begin{equation}
	\frac{p^2+pp'+p^{\prime 2} +\chi_-^2 + \chi_+^2 -3 \tilde{q} (p+p') +2\tilde{q}^2}{2(p-p')^2} \left[ 2(\chi_- - \chi_+)^2 +2 p p' \right]^2 \ .
\end{equation}
In contrast, the other two contributions require computing a different one-loop integral
\begin{equation}
	I^{(2)}=\int{\frac{d\Omega d\rho}{2\pi} \frac{\left[ V_4^{(0)} (p,p',\rho,p+p'-\rho) + \frac{V_4^{(1)} (p,p',\rho,p+p'-\rho)}{\lambda} \right]^2}{[\Omega - \omega(\rho) +i\epsilon][\omega(p) + \omega(p') -\Omega -\omega(p+p'-\rho) +i\epsilon]}} \ .
\end{equation}
The integral over $\Omega$ can be performed as in the previous case, giving us
\begin{align}
	\lambda I^{(2)} &= \lambda \int{d\rho \frac{\left[ V_4^{(0)} (p,p',\rho,p+p'-\rho) + \frac{V_4^{(1)} (p,p',\rho,p+p'-\rho)}{\lambda} \right]^2}{\omega(p) + \omega(p') -\omega(\rho) -\omega(p+p'-\rho) +2 i\epsilon}} \notag \\
	&= \lambda I^{(1)} + \int{d\rho \frac{2 V_4^{(0)} (p,p',\rho,p+p'-\rho) V_4^{(1)} (p,p',\rho,p+p'-\rho)}{\omega^{(0)}(p) + \omega^{(0)}(p') -\omega^{(0)}(\rho) -\omega^{(0)} (p+p'-\rho) +2 i\epsilon}} \notag \\
	&+ \int{d\rho \frac{\left[ V_4^{(0)} (p,p',\rho,p+p'-\rho) \right]^2 \left[\omega^{(1)}(p) + \omega^{(1)}(p') -\omega^{(1)}(\rho) -\omega^{(1)} (p+p'-\rho) \right] }{\left[ \omega^{(0)}(p) + \omega^{(0)}(p') -\omega^{(0)}(\rho) -\omega^{(0)} (p+p'-\rho) +2 i\epsilon \right]^2}} +\mathcal{O} \left( \frac{1}{\lambda} \right) \ ,
\end{align}
where we have separated the terms in the dispersion relation as $\omega (p) = \omega^{(0)} (p)+ \omega^{(1)} (p)\lambda^{-1} + \mathcal{O} \left( \lambda^{-2} \right)$. The first integral is the one we solved above for the 1-loop at leading order. The second one is a similar integral with a different numerator, thus it can be solved similarly
\begin{align}
	&\int{d\rho \frac{2 V_4^{(0)} (p,p',\rho,p+p'-\rho) V_4^{(1)} (p,p',\rho,p+p'-\rho)}{\omega^{(0)}(p) + \omega^{(0)}(p') -\omega^{(0)}(\rho) -\omega^{(0)} (p+p'-\rho) +2 i\epsilon}}=\frac{2(\chi_- - \chi_+)^2 +2 p p'}{p-p'} \\
	&\times \bigg( 2 (\chi_+ \chi_- + p p') (\chi_- - \chi_+)^2 + p p' (p^2+p^{\prime 2}) + \tilde{q} (p+p') [2p p' + (\chi_- - \chi_+)^2] +2 p p'\left[\chi_+ \chi_- +\tilde{q}^2 \right] \bigg) \ . \notag
\end{align}
In contrast, the third contribution has second order poles instead. Luckily, these are placed at the same points as the poles of the other two integrals, so the computation is not much different. Nevertheless, there is a way around this computation. Let us consider instead the integral without expanding the energy
\begin{equation}
	\int{d\rho \frac{\left[ V_4^{(0)} (p,p',\rho,p+p'-\rho) \right]^2}{\omega(p) + \omega(p') -\omega(\rho) -\omega(p+p'-\rho) +2 i\epsilon}} \ ,
\end{equation}
this integral can also be computed by residues. The integrand has simple poles located at $\rho=p$ and $\rho=p'$. If we choose to close the contour to pick only the first of them, we get
\begin{equation}
	\int{d\rho \frac{\left[ V_4^{(0)} (p,p',\rho,p+p'-\rho) \right]^2}{\omega(p) + \omega(p') -\omega(\rho) -\omega(p+p'-\rho) +2 i\epsilon}}=\frac{\left[ V_4^{(0)} (p,p',p,p') \right]^2}{\left. \frac{\partial \omega (q)}{\partial q} \right|_{q\rightarrow p} - \left. \frac{\partial \omega (q)}{\partial q} \right|_{q\rightarrow p'}} \ .
\end{equation}
Notice that the denominator is just the same as the kinematic factor. We can extract the third contribution to $I^{(2)}$ by isolating the $\lambda^{-1}$ term of this result.

Thus, the $1$-loop contribution at next-to-leading order in $\lambda$ can be written as
\begin{equation}
	\lambda S_{\text{1-loop}}^{(1)}=(-i)^2 \left[ \left( K^{(0)}\right)^2 V^{(0)}_4(p,p',p,p') V^{(1)}_4(p,p',p,p') + 2 K^{(0)} K^{(1)} \left( V^{(0)}_4(p,p',p,p') \right)^2 \right] \ .
\end{equation}


\subsubsection{A comment on the all-order vertex}


Although we can compute more terms of the Lagrangian and use them to compute further corrections in $\lambda$ of the $S$-matrix, we would like to instead take a look at the trick proposed in \cite{Roiban:2006yc} to obtain the $S$-matrix  at all orders in $\lambda$ for the undeformed $AdS_5 \times S^5$.

The idea goes as follows: if one substitutes the leading-order vertex one gets from the Lagrangian, $p p' + k k'$, by the vertex $\frac{p p'}{e(p) e(p')}+\frac{k k'}{e(k) e(k')}$ (here $e(p)=\sqrt{1+p^2}$ is the all-order energy of the excitations in that setting), the sum of all the $n$-loop contributions reconstructs the BDS $S$-matrix instead of the Heisenberg $S$-matrix. Although the process gives the correct result, the computation is not free of ambiguities. In particular, the next-to-leading order of the vertex is not the one obtained from the expansion of the Lagrangian. Nevertheless, the mismatch between them is an off-shell quantity, so it can be attributed to a field redefinition.

Inspired by this result, one may be tempted to conjecture that we can get an all-order $S$-matrix in the deformed case by using the vertex
\begin{equation}
	-\langle k , k' | \tilde{V}_4 | p , p'\rangle=-\tilde{V}_4(p,p',k,k') =\frac{(\chi_- - \chi_+)^2 +p p'}{\hat{\Omega}_- (p)\hat{\Omega}_- (p')} +\frac{(\chi_- - \chi_+)^2+ k k'}{\hat{\Omega}_- (k)\hat{\Omega}_- (k')} \ ,
\end{equation}
where $\hat{\Omega}_- (p)$ is the expression obtained after performing the appropriate rescaling of the parameters to the exact dispersion relation $\Omega_-(p)$ defined in equation (\ref{ExactEnergy}). However, this simple substitution does not work in the deformed case. We can see that by expanding the proposed vertex on-shell (i.e., if we set $k=p$ and $k'=p'$) at large tension
\begin{align}
	&\frac{-\tilde{V}_4(p,p',p,p')}{2} =\frac{(\chi_- - \chi_+)^2 +p p'}{\hat{\Omega}_- (p)\hat{\Omega}_- (p')} \approx (\chi_- - \chi_+)^2 +p p' -[(\chi_- - \chi_+)^2 +p p'] \frac{\omega(p) + \omega (p')}{2\lambda} + \mathcal{O} \left( \lambda^{-2} \right) \notag \\
	&=  (\chi_- - \chi_+)^2 +p p' -\frac{[(\chi_- - \chi_+)^2 +p p'] [2 (\chi_- - \chi_+)^2 +p^2 + p^{\prime 2} -2\tilde{q} (p+p')]}{2\lambda} + \mathcal{O} \left( \lambda^{-2} \right) \ .
\end{align}
Compare this expression with the one we have obtained from the Lagrangian
\begin{align}
	&\frac{-1}{2}\left[ V^{(0)}_4(p,p',p,p')+\frac{V^{(1)}_4(p,p',p,p')}{\lambda} + \mathcal{O} \left( \lambda^{-2} \right) \right] = (\chi_- - \chi_+)^2 +p p' -\frac{2 (\chi_+ \chi_- + p p') (\chi_- - \chi_+)^2 }{2\lambda}  \notag \\
	& -\frac{ p p' (p^2+p^{\prime 2}) + \tilde{q} (p+p') [2p p' + (\chi_- - \chi_+)^2] +2 p p'\left[\chi_+ \chi_- +\tilde{q}^2 \right]}{2\lambda} + \mathcal{O} \left( \lambda^{-2} \right) \ .
\end{align}
These two quantities only match when $\chi_\pm=\tilde{q}=0$. Among other differences, the mismatch is most evident in the $\tilde{q}^2/\lambda$ term, which vanishes for the first expansion but not for the second one. The two expressions do not match on-shell, meaning that the modification of vertex we have proposed do not correctly capture the form of the deformed Lagrangian. Thus, the simple trick from \cite{Roiban:2006yc} cannot be generalised straightforwardly to this deformed setting.

\section{Conclusions}

In this article we studied the Landau-Lifshitz limit of the non-linear sigma model associated with the three-parameter deformation of the $\mathbb{R}\times S^3\subset AdS_3\times S^3 \times T^4$ background. We showed that the dispersion relation of near-BMN excitations associated to this action is consistent with the expression computed from uniform light-cone gauge quantisation \cite{3deformedSMatrix}. We also constructed an all-loop $S$-matrix at leading order in the string tension $\lambda$. Our result becomes the Heisenberg $S$-matrix at zero deformation, as expected from an $SU(2)$-like sector. In addition, the tree-level contribution to this $S$-matrix agrees with the low-energy of the tree-level contribution computed in \cite{3deformedSMatrix}. We have also computed the tree-level contribution at next-to-leading order in $\lambda$.  We found that the $S$-matrix only depends on the combination of the deformation parameters $\chi_- - \chi_+$ at leading order, with $\tilde{q}$ and the combination $\chi_+ \chi_+$ appearing in the first correction in $\lambda$. In addition, we showed that the modification of the diagrammatic procedure proposed in \cite{Roiban:2006yc} does not extend immediately to our setting.

One immediate question that we would like to investigate is if there exists a generalisation of the method from \cite{Roiban:2006yc} to this deformed background. The analysis of the low-energy limit of the $S$-matrix of the $\eta$-deformed $AdS_3\times S^3 \times T^4$ space might shed some light on this question. However, we should not discard the possibility that the modification actually works and that the mismatch was created by our procedure. We are substituting the Virasoro constraints on the action instead of the equations of motion. This step has the risk of giving the wrong Lagrangian, although we have some reasons to think that this is not the case. First, we know that this does not happen at leading and next-to-leading order in the undeformed case \cite{Kruczenski:2004kw}. In addition, we have checked that the leading order of our Lagrangian gives rise to the correct equations of motion. However, explicitly checking the equations of motion at next-to-leading order proves to be cumbersome due to the length of the contribution and the field redefinitions involved. Despite that, the fact that we have obtained the correct expression for the dispersion relation can be considered a good circumstantial evidence that there is no issue with the substitution also at next-to-leading order. Nevertheless, it would be interesting to check if that is the case.

In addition, it would be interesting to study the generalisation of our Lagrangian to the full deformed $AdS_3\times S^3 \times T^4$ space. The procedure is well understood for the case of IIB superstring in $AdS_5  \times S^5$, see e.g. \cite{Stefanski:2007dp}, and we expect our case to be relatively similar. Sadly, including the fermionic degrees of freedoms requires us to know the RR flux associated to this background. As we mentioned, the problem of finding the correct flux for which the background is a solution of the supergravity equations is not a trivial task. Nevertheless, we can still analyse just the bosonic part of the action, as it does not depend on the RR flux. A different generalisation would be to consider higher orders in $1/J$ and study the three-body $S$-matrix. It would be interesting to repeat the computation from \cite{Melikyan:2008cy} and check if the three-particle scattering matrix still factorises.

We would also like to make some comments about our choice of Kalb-Ramond field. Our choice was motivated by matching with the choice for the mixed-flux deformation from \cite{Stepanchuk}. Their choice was the only one that made the dispersion relation of the dyonic giant magnon finite. Here we have commented that our choice made the leading order of the effective field theory free of a total derivative at order $J^2$. However, we are not in a position to claim that such a condition does not suffer a deformation. Due to the small deformation limit, we can only claim that such deformation has to be at least quadratic in $\chi_\pm$ or linear in $q$, as there is no total derivative term also at next-to-leading order in $\lambda$. A more detailed study of classical solutions would be necessary to check if there is no deformation.

\section{Acknowledgements}

We are grateful to Rafael Hernández, Roberto Ruiz, Bogdan Stefanski, Alessandro Torrielli and Kentaroh Yoshida for reading the manuscript and providing very useful comments. LW is funded by a University of Surrey Doctoral College Studentship Award. This work is supported by the EPSRC-SFI grant EP/S020888/1 {\it Solving Spins and Strings}. 

No data beyond those presented and cited in this work are needed to validate this study.

\appendix

\section{Next-to-leading order Landau-Lifshitz action} \label{subleading}

In this appendix, we will present some details of the computation of the next-to-leading order contribution to the Lagrangian from the expansion at large energy of (\ref{beforeexpansion}). In particular, we want to emphasise some peculiarities that we do not find when computing the leading order contribution.

We approach the problem in a similar way as before: We start with the Lagrangian~(\ref{beforeexpansion}) and its Virasoro constraints in terms of the coordinates $r, \alpha, \beta$ and perform the same rescaling of the parameters as before. We expand the resulting expressions for the Lagrangian and the Virasoro constraint that involves crossed derivatives up to the order $\kappa^{-2}$. Regarding the remaining Virasoro constraint, equation~(\ref{vivi}) is enough for our purposes. This is the case because the leading order of the Lagrangian does not depend on $\dot{\alpha}$ except for the total derivative term, which can be disregarded right away. The Virasoro constraint that involves crossed derivative takes the form
\begin{align*}
0 + O\left(\frac{1}{\kappa^3 }\right) &= \kappa  \left[\left(1-2 r^2\right) {\beta^\prime}+\alpha^\prime \right]+\frac{1}{\kappa} \left\{ \alpha^\prime \left[ (\chi_-^2 - 4 \chi_- \chi_+ + \chi_+^2) r^2 (r^2-1) + \dot{\alpha} + (1-2 r^2) \dot{\beta} \right] \vphantom{\frac{r^\prime \dot{r}}{r^2-1}} \right. \\
&+ \left. \left[-(\chi_-^2-\chi_+^2) r^2 (r^2-1)+ (1-2 r^2) \dot{\alpha} + \dot{\beta} \right] \beta^\prime - \frac{r^\prime \dot{r}}{r^2-1} \right\} \ ,
\end{align*}
We can solve this equation for $\alpha'$ as a series in $\kappa$ and substitute it back into the Lagrangian, together with eq.~(\ref{vivi}), to get rid of the coordinate $\alpha$. We should stress here that $\alpha'$ has a contribution at order $\kappa^{-2}$, so the leading order of the Lagrangian generates a contribution to the next-to-leading order through this substitution.

The next point we have to address is the dependence of the next-to-leading order on the square of the generalised velocities of $\beta$ and $r$. This dependence would complicate our Lagrangian, so we would like to get rid of those contributions. This can be done through a field redefinitions if we follow the procedure detailed in appendix A of \cite{Kruczenski:2004kw}. If we perform a field redefinition of our coordinates of the form $\beta\rightarrow \beta+\frac{\tilde{\beta}}{\kappa^2} +\dots$ and $r\rightarrow r+\frac{\tilde{r}}{\kappa^2} +\dots$, the expansion of the Lagrangian in powers of $\kappa$ becomes
\begin{equation}
	\mathcal{L}=\mathcal{L}^{(0)}+\frac{\mathcal{L}^{(2)}}{\kappa^2} +\dots \rightarrow \mathcal{L}=\mathcal{L}^{(0)}+\frac{\mathcal{L}^{(2)}+\tilde{\beta}\delta_\beta \mathcal{L}^{(0)}+\tilde{r}\delta_r \mathcal{L}^{(0)} }{\kappa^2} +\dots \ ,
\end{equation}
where $\delta \mathcal{L}$ represent the variational derivative. The cornerstone of this procedure is the fact that our Lagrangian depends linearly on the generalised velocities of the fields. This implies that the variational derivatives have to depend (at most) linearly on the generalised velocities.\footnote{One can see that a Lagrangian of the form $\mathcal{L} = A(r, \beta) \dot{\beta} + B(r, \beta) \dot{r} + V (r,\beta)$ has $\delta_\beta \mathcal{L} = \partial_r A(r,\beta) \dot{r} + \partial_{\beta} A(r,\beta) \dot{\beta} + \partial_{\beta} V $ and $\delta_r \mathcal{L} = \partial_r B(r,\beta) \dot{r} + \partial_{\beta} B(r,\beta) \dot{\beta} + \partial_r V $.} In our particular case we have
\begin{align*}
\frac{1}{8}\delta_\beta \mathcal{L}^{(0)} &= \left[\left(2-4 r^2\right) \beta^\prime -\tilde{q} \right] r r^\prime +\left(1-r^2\right) r^2 \beta^{\prime \prime}+r  \dot{r} \ , \\
\delta_r \mathcal{L}^{(0)} &= r \left\{-2 \left[(\chi_- -\chi_+ )^2-4 \tilde{q} \beta^\prime +4 {\beta^\prime}^2\right]+\frac{2\left(r^\prime\right)^2}{\left(r^2-1\right)^2}-8 \dot{\beta} \right\}+4 r^3 \left((\chi_- -\chi_+ )^2+4 {\beta^\prime}^2\right)-\frac{2 r^{\prime \prime}}{r^2-1} \ .
\end{align*}
We can see that they are linear in $\dot{r}$ and $\dot{\beta}$ respectively. Thus, one just has to find the form of the functions $\tilde{\beta}$ and $\tilde{r}$ that make the next-to-leading order of the Lagrangian independent of the generalised velocities. In our case we find
\begin{align}
	\tilde{\beta}&=\frac{r^\prime \left[\tilde{q}+\left(4 r^2-2\right) \beta^\prime \right]+r \left(r^2-1\right) \beta^{\prime \prime}}{8 r \left(r^2-1\right)} + \frac{\dot{r}}{8 r (r^2-1)} \ , \notag \\
	8 \tilde{r}&=r''-r(1-r^2) \left[ ((1 + 2 r^2) (\chi_+ + \chi_-)^2 - 4\chi_- (\chi_- + 2 \chi_+ r^2) -   4 \tilde{q} \beta ' + 4(1 - 2 r^2) \beta '' \right] \notag \\
	&+ \frac{r r^{\prime 2}}{(1-r^2) } +4 r(1-r^2)\dot{\beta}\ .
\end{align}
\pagebreak

Substituting these field transformations, we arrive at
\begin{align}
	\mathcal{L}^{(2)}&=-\frac{r^{\prime 4}}{4 (1-r^2)^3} +\frac{r^{\prime 2} \left[ (1 - r^2) (\chi_+^2 r^2 (2 - r^2) - 2 \chi_+ \chi_- r^2 (1 - r^2) + \chi_-^2 (2 - r^4)) + r r'' \right]}{2 (1-r^2)^2} \notag \\
		& +\frac{4\tilde{q}^2 r^{\prime 2}+(r'')^2}{4(1-r^2)}+\frac{1}{4} (\chi_- - \chi_+)^2 r^2 (1 - r^2) [(\chi_- - \chi_+)^2 (1 + r^2 - r^4) -  4 \chi_- \chi_+]  \notag \\
		&+ \frac{1}{2} (\chi_-^2 - \chi_+^2) r r'' +\dots \ ,
\end{align}
where we have just reproduced the part that depends only on $r$ for brevity. Similarly to the leading order contribution, it is more useful to express the Lagrangian in terms of a complex field. Using again the parameterisation $\phi=\sqrt{1-r^2} e^{2 i \beta}$, performing the rescaling~(\ref{secondscaling}), and taking the limit of large $J$ we arrive at the expression (\ref{SecondL}).

\end{document}